**Effects of Pressure on Electron Transport and Atomic Structure of Manganites:**

**Low to High Pressure Regimes**

Congwu Cui[a], Trevor A. Tyson[a], Zhong Zhong[b], Jeremy P. Carlo[a] and Yuhai Qin[a]

[a] Physics Deportment, New Jersey Institute of Technology, Newark, NJ 07102

[b] National Synchrotron Light Source, Brookhaven National Laboratory, Upton, NY

11973

## Abstract

The pressure dependence of the resistivity and structure of $La_{0.60}Y_{0.07}Ca_{0.33}MnO_3$ has been explored in the pressure range from 1 atm to ~7 GPa. The metal to insulator transition temperature ($T_{MI}$) was found to reach a maximum and the resistivity achieves a minimum at ~3.8 GPa. Beyond this pressure, $T_{MI}$ is reduced with a concomitant increase in the resistivity. Structural measurements at room temperature show that at low pressure (below 2 GPa) the Mn-O bond lengths are compressed. Between ~2 and ~4 GPa, a pressure induced enhancement of the Jahn-Teller (JT) distortion occurs in parallel with an increase in Mn-O1-Mn bond angle to ~180°. Above ~4 GPa, the Mn-O1-Mn bond angle is reduced while the JT distortion appears to remain unchanged. The resistivity above $T_{MI}$ is well modeled by variable range hopping. The pressure dependence of the localization length follows the behavior of $T_{MI}$.

# I. INTRODUCTION

In the $La_{1-x}A_xMnO_3$ (A = Ca, Sr) system, when x is in the range of 0.2~0.5, there is a metal-insulator transition (MIT) with increasing temperature and the Curie temperature $T_c$ coincides with the MIT temperature $T_{MI}$.[1] This can be explained qualitatively by the double exchange model (DE).[2] But the predicted resistivity[2] is much lower than that from experimental measurements. Millis *et al.*[3] argued that DE alone cannot explain the resistivity in these systems (also called colossal magnetoresistance (CMR) materials) and that local lattice distortions, specifically Jahn-Teller (JT) type lattice distortions of the $MnO_6$ octahedra, should be considered. Due to JT distortions (JTD), the degenerate $Mn^{3+}$ $e_g$ orbital splits, thus lowering the energy of the occupied orbital and localizing the state. Because of the subtle balance and complicated interactions among the charge, spin and lattice structure (symmetry and local atomic structure), many experimental parameters, such as the average A-site radius, magnetic fields, high pressure and photons, can affect the transport properties and cause changes in magnetic and/or structural order.

In the cubic perovskite structure $ABO_3$, due to the radius mismatch of the A and B site atoms, structural distortion is induced. By chemical substitution at the A site, not only the number of electrons in 3d band of Mn and the lattice parameters but also the Mn-O bond length and Mn-O-Mn bond angle are changed.[1] The local distortion can also be changed with different doping levels. For systems of $Ln_{1-x}A_xMnO_3$ (Ln = La, Pr, Nd, Sm, Y etc; A = Ca, Sr, Ba, Pb etc.), the magnetic, electronic and structural properties have been investigated by changing the doping elements and level x, resulting in detailed phase diagrams.[2, 4]

Unlike internal (or chemical) pressure induced by chemical doping (which may change both the Mn valence and structure), hydrostatic pressure is a "clean" method to change the long range and local structure in the CMR materials in a continuously tunable way. High pressure has been found to stabilize the rhombohedral phase in the $La_{1-x}Sr_xMnO_3$ system (x=0.12~0.18)[5, 6] and $La_{0.8}Ba_{0.2}MnO_3$.[7] In the low-pressure range, the effects of pressure on the manganites can be accounted by DE theory. Generally, it is believed that pressure compresses the lattice constants, increases the Mn-O-Mn bond angle, makes the unit cell more cubic and hence reduces the local distortion of the $MnO_6$ octahedra and electron-lattice coupling. As a result, the overlap of the $Mn^{3+}$ $e_g$ orbital and $O^{2-}$ 2p orbital is increased- thus enhancing the electron hopping rate. Indeed, for many systems with paramagnetic insulating (PMI) to ferromagnetic metallic (FMM) phase transitions, $T_c$ increases almost linearly with pressure in the pressure range below 2 GPa[6, 8, 9, 10, 11, 12, 13, 14, 15, 16], with few exceptions.[17] But the pressure effect on $T_c$ is larger than that predicted by band theory. This implies that the electron-phonon coupling is also reduced by pressure.[12] The sensitivity of $T_c$ to pressure, $dT_c/dP$, depends on the doping level or the A-site average radius $<r_A>$.[17, 13] This is due to the fact that manganites with small $<r_A>$ have larger local distortions and hence can theoretically go through a larger degree of ordering with pressure.[13] In $La_{1-x}Sr_xMnO_3$ (x=0.12-0.18) pressure was also found to be able to destabilize ordered JT polarons, to enhance electron hoping and extend the FMM state to lower temperature; in comparison, magnetic fields have negligible effect on these combined parameters which suggest that spin ordering plays a minor role in this system.[6]

LaMnO_3, the prototypical parent compound, is an A-type antiferromagnetic insulator with highly coherent static Jahn-Teller distortions (with octahedral bond distances of 1.907 Å, 1.968 Å and 2.178 Å).[18] Under pressure it first undergoes a transition from localized electron to band antiferromagnetism at ~0.7 GPa.[19] With further pressure increase the $MnO_6$ octahedra are nearly isotropically compressed and the Jahn-Teller distortion remains stable up to ~ 7 GPa. In this range, pressure decreases the orthorhombic distortion by reducing the average tilt angle of $MnO_6$ octahedra. Consequently, the magnetic ordering temperature and electronic bandwidth are increased. Above 7 GPa the compound possibly undergoes a transition to a metallic-like phase.[20] In the manganites, Jahn-Teller distortions (static and dynamic) play an important role.[21] When crossing into the FM phase both coherent and incoherent distortions are abruptly reduced. The coherence state of distortions may be affected by high pressure and doping.[22] Also, the electron-phonon interaction can be affected by pressure by modifying the "stiffness" of the phonons and the distortion modes by enhancing the Q3 mode and suppressing the Q2 mode.[23]

By comparing the effects of chemical doping and pressure in the range below ~2 GPa (the upper limit of the traditional clamp pressure cells), it has become generally accepted that the effects of hydrostatic pressure is equivalent to that of chemical doping. Hwang *et al.*[24] systematically studied the effects of external hydrostatic pressure and internal chemical pressure on the properties of CMR and found that up to ~2 GPa the effect of hydrostatic pressure can be mapped onto the average radius of the A site atoms with a conversion factor of $3.75 \times 10^{-4}$ Å/kbar.

There have been some indirect indications that, for pressures above 2 GPa, the behavior of CMR oxides may be different from that observed in the low-pressure measurements. Raman scattering result by Congeduti *et al.*[25] on $La_{0.75}Ca_{0.25}MnO_3$ indicated that above 7.5 GPa, high pressure induces a new phase other than the predicated metallic phase. The abrupt phonon frequency change and strong phonon broadening suggest a charge-lattice interaction strengthened by the lattice compression. Meneghini *et al*.'s[26] results revealed that in addition to the general unit cell contraction, pressures above 6-7GPa cause the $MnO_6$ octahedra to become more distorted by splitting the two almost identical in-plane Mn-O bond lengths and produce a longer range static/dynamic JTD. However, because of the high transition temperature of this material, only a limited study of the changes in transport with temperature could be observed. We have studied the system $La_{0.60}Y_{0.07}Ca_{0.33}MnO_3$ with a transition temperature that enables the observation of shifts in $T_{MI}$ over a broad range of pressures.

Here we report our results of electric transport and structure of $La_{0.60}Y_{0.07}Ca_{0.33}MnO_3$ under high-pressures up to ~7 GPa. This compound has a very high magnetoresistance of ~10000% at 6 T.[27] Its Curie temperature $T_c$ and MIT temperature $T_{MI}$ coincide at ~150 K. Its magnetotransport properties suggest strong electron-lattice and spin-lattice coupling.[28] For pressures up to ~0.8 GPa, $T_c$, $T_{MI}$ and the linear thermal expansion coefficient peak coincide and are linear functions of pressure.[8] Although this material has been extensively studied, its properties under high pressure above 2 GPa were still unexplored. We found that below P*~ 3.8 GPa, high pressure increases $T_{MI}$ and suppresses resistivity. But above P*, $T_{MI}$ decreases and the resistivity increases quickly with pressure. The resistivity in the measured temperature range of liquid nitrogen to room temperature follows the same manner. This possibly suggests that high pressure causes a change in the crystal structure (local or long range). Hence, high-pressure X-ray diffraction measurements were performed to determine the structural evolution under high pressure.

We found that at P* pressure induced a structural transformation within the $MnO_6$ octahedra to a highly JT distorted state. Above P*, with increasing pressure the $MnO_6$ octahedra continue to tilt.

## II.  EXPERIMENTAL METHODS

Samples of $La_{0.60}Y_{0.07}Ca_{0.33}MnO_3$ were prepared by solid-state reaction with multiple cycles of grinding and calcination at a temperature of 1200 °C in air.  The resulting powder was then pressed into pellets and annealed in air at 1300 °C for 12 hours and slowly cooled down to room temperature at a rate 1°C/min.

The X-ray diffraction pattern taken at room temperature with a Rigaku X-ray diffractometer with a Cu sealed tube showed that the samples are in a single crystallographic phase (Fig. 1). The structure was refined to Pbnm symmetry using the Rietveld method. The refined lattice constants are: a = 5.45810(6) Å, b = 5.45149(7) Å, c = 7.69806(11) Å. The sample was also characterized by magnetization measurements (Inset of Fig. 1). The magnetic moment at 5 K in a 10 kOe magnetic field is $3.66\mu_B$ which compares well with the theoretical estimate of $3.67\mu_B$. The Curie temperature is defined as the edge, the maximum of the first order derivative of the magnetization *vs.* temperature curve. The $T_c$ extracted in this way is 150 ± 2.5 K - consistent with the metal-insulator transition temperature $T_{MI}$ (149.8 ± 1.0 K), the temperature at the resistivity peak.  (We note also that magnetization measurements in a low field of 10 Oe yield a $T_c$ value of 145 ± 2.5 K)

High-pressure transport measurements were carried out with a diamond anvil cell. The culet size of the diamond anvils is 800 μm. Samples for high pressure resistivity measurements were cut from a pellet, polished to a sheet ~60 μm thick and then cut to small pieces of 100~200 μm dimension. Four gold wires were glued to the four corners of the sample with silver paste. Then the sample was heat treated at ~80 °C for several hours for the silver paste to cure. The stainless steel gaskets and the wall of the sample chamber were coated with a thin layer of 1:1 Stycast 1266 epoxy and $Al_2O_3$ powder mixture for electrical insulation. Fluorinert FC-77 was used as the pressure medium. Two or three ruby chips were placed around the sample in the gasket hole for pressure calibration. For a given pressure setting, at different temperatures (20-40 K steps) and multiple positions near the sample the ruby fluorescence shifts were measured. The sample pressure was then calculated from the average and the errors were estimated using the standard deviations of the ~8-20 pressure measurements. The resistivity was measured using the Van Der Pauw four-point method. Since rapid cool down was less stable, data were collected only while warming up.

High-pressure X-ray diffraction experiments were performed at beamline X17B1 at the NSLS, Brookhaven National Laboratory, in transmission mode through the two diamond anvils and a CCD (Mar, 2048 × 2048 pixels with 79 micron resolution) was used to obtain the diffraction patterns. The images were converted to intensity *vs.* 2θ by integrating around the rings of the powder pattern using the program FIT2D. The wavelength of the X-rays was 0.185 Å. The intensity, energy resolution, and the in-plane divergence of the x-rays are $10^{11}$ photons/s.mm$^2$, $10^{-4}$ (dE/E) and 0.1 milli-radians, respectively.[29, 30] The x-rays were sagittally-focused[30] from a width of 20 mm to 0.4 mm to increase the x-ray intensity on the sample and then apertured to minimize background scattering by the gasket material. The data were collected from four samples and care was taken to avoid gasket deformation which can modify the background from gasket contribution to the diffraction pattern. The pressure medium used for X-ray diffraction is 4:1 methanol-ethanol and 2-3 ruby chips were used for pressure calibration (as in the transport measurements). For these measurements, the pressure is hydrostatic up to at least 10 GPa, the only errors are time dependent changes in pressures. At all the measured pressures, the maximum time dependent change is ~0.1 GPa. All diffraction data were refined by the Rietveld method using the program Rietica. Figure. 2 shows two typical sets of data at ambient pressure and 5.9 GPa. The shaded regions (not used in the fits) correspond to diffraction from the steel gasket and random narrow noise spikes.

## III. RESULTS AND DISCUSSION

### A. Transport Measurements

The resistance of the sample as a function of temperature, under pressures up to ~7 GPa is shown in Fig. 3. Fig. 4(a) is the pressure dependence of $T_{MI}$. It is apparent that $T_{MI}$ increases first, saturates and then quickly drops with increasing pressure. At ambient pressure, $T_c$ and $T_{MI}$ coincide. In the same material, it was reported that $T_c$ and $T_{MI}$ still coincide under pressure up to ~0.8 GPa.[8] In the parent compound $La_{0.67}Ca_{0.33}MnO_3$, Tc and $T_{MI}$ coincide up to at least 1.6 GPa.[17] We are unaware of results on the coincidence of $T_c$ and $T_{MI}$ beyond this pressure range. However, it has been reported that the substitution of La atom with Gd and Y leads to a separation between $T_c$ and $T_{MI}$.[31, 32] Hence, in the higher pressure range this question is still open. In this paper we discuss shifts in $T_{MI}$ and leave open the question of shifts in Tc at pressure above 1.6 GPa for future work.

In Fig. 4(a) the data for $T_{MI}$ *vs.* P is fitted with a third order polynomial. The $dT_c/dP$ (or $dT_{MI}/dP$) near ambient pressure determined with it is 22 ± 4 K/GPa. It is consistent with the 26 ± 2 K/GPa value reported on the same material by Arnold *et al.*[8]

Another noticeable feature about the resistivity data at different pressures is the peak width. The peak width is defined as full width at half maximum (FWHM). With pressure increase, the peak is dramatically broadened (Fig. 4(b)). This may originate from non-hydrostatic pressure conditions. By placing multiple ruby chips in the cell, we found that the pressure difference around the sample increases with pressure which may imply that the pressure medium freezes more easily at higher pressure. Because the size of the ruby chips is quite small (< 10 μm), the fluorescence doublet still separate very well except that the peaks are only slightly broadened. The largest difference of the pressure observed around the sample is ~0.5. The pressure was also found to decrease with temperature increase. The higher pressure, the larger this pressure changing is. The overall variations in the pressure in the sample space are indicated as error bars in the related figures. The variation of pressure around the sample and with temperature does not explain the peak broadening. Apparently, the main reason for the peak broadening may be that the material is becoming insulating with pressure increase so that the peak is suppressed and disappears.

The conductivity in the whole temperature range changes in the same behavior as $T_{MI}$, the only difference is that the resistivity in the metallic region changes faster than in the paramagnetic insulating region. The $T_c$ (and $T_{MI}$) of the parent compound

La$_{0.7}$Ca$_{0.3}$MnO$_3$ is ~270 K. Under high pressure, the T$_{MI}$ of La$_{0.60}$Y$_{0.07}$Ca$_{0.33}$MnO$_3$ does not reach 270 K but saturates far below at ~215 K and then quickly decreases with increasing pressure.

It was reported that in a similar compound La$_{0.6}$Y$_{0.1}$Ca$_{0.3}$MnO$_3$ the resistivity in the paramagnetic phase follows a variable range hopping (VRH) model in which the resistivity behaves as ~exp(T$_0$/T)$^{1/4}$.[33] Compared with the adiabatic and nonadiabatic polaron models, the measurements here are consistent with the VRH behavior (Fig. 5(a)). The localization length was estimated according to Viret *et al.*'s[34] magnetic localization theory, which suggests that the mechanism of MIT is localization associated with magnetic disorder. Based on this theory, the localization length $\xi$ can be expressed as:

$$\xi^3 = \frac{120U_m(1 - <\cos\theta_{ij}>)v}{kT_0 g} \tag{1}$$

where U$_m$ (= 3J$_H$/2) is the Hund's rule coupling strength; $\theta_{ij}$ is the angle between the two neighbor spins; v is the lattice volume per manganese ion; g is the probability that an unoccupied manganese orbital can actually accept an electron, which reflects the dynamic JT effect, only when the receiving site is not distorted or properly distorted can electron hopping happen. In the above equation, the localization length is the function of both the Mn-O-Mn bond angle and the dynamical JTD.

The localization length extracted according to this model is shown in Fig. 5(b). The maximum of localization length at ~P* is ~0.21 nm. This is the order of the Mn-O bond length. The corresponding hopping distance is ~1.35 nm which is several unit cells. It is noticeable that this is also the size of the magnetic clusters Sun *et al.*[35] reported.

Polaron models are also extensively used to explain the transport behavior of manganites. It was reported that the variable range hopping of small polarons can also leads to $\ln(\rho) \propto T^{-1/4}$ behavior.[36] Kapusta *et al.*[37] suggested that the magnetic correlations in systems of $(La_{1-x}A_x)MnO_3$ (A = Ca, Sr) be possibly due to magnetic polarons. With temperature decrease there is a transition from small-polaron-dominated PMI regime to a large-polaron-dominated FMM regime.[38] Röder *et al.*[39] reported that above $T_c$ the small magnetopolaron due to the JT coupling, which involves about 4 lattice sites, comprises a localized charge surrounded by a spin cloud on nearest neighbors. Small angle neutron scattering measurements on $La_{2/3}Ca_{1/3}MnO_3$ found that the magnetic polarons have dimensions of the order of ~1.2 nm above $T_c$ and that high magnetic fields enhance the correlation length significantly.[40]

Despite the difference between the models, magnetic localization and the polaron formation depend critically on the local structure. The distortion of local structure, such as static and dynamic Jahn-Teller distortion and/or rotation of the $MnO_6$ octahedra, play an important role on the transport behavior.

### B. Structural Measurements

To understand the high-pressure resistivity results, high-pressure X-ray diffraction measurements were performed. The data were refined with the Rietveld method on the basis of the 1 atm Pbnm space group. The pressure dependence of unit cell volume is shown in Fig. 6(a). In the measured pressure range, it is monotonically compressed. In Fig. 6(b) and 6(c) are the Mn-O bond length and Mn-O-Mn bond angle pressure dependence, respectively. Below ~2 GPa, all three Mn-O bonds are compressed and the bond angles have no obvious change. This may explain why the $T_{MI}$ and resistivity behave according to the DE theory: The pressure compresses the Mn-O bonds to increase the $Mn^{3+}$ $e_g$ band and $O^{2-}$ 2p band overlap, enhancing the hoping integral. From ~2 to ~3 GPa, there is a local structure transformation similar to that in $La_{0.75}Ca_{0.25}MnO_3$.[26] The splitting of the two in-plane Mn-O2 bonds increases. The Mn-O1-Mn bond angle increases by about ~20° while the Mn-O2-Mn bond angle seems only decrease slightly. In the meantime the coherent Jahn-Teller distortion, defined as the deviation of Mn-O bonds from average, increases abruptly (Fig. 6(d)). Meneghini *et al.*[26] suggested a transition to a coherent local and/or dynamical Jahn-Teller distortion. This can partly explain why the $T_c$ increase and resistivity decrease are halted at high pressure. With enhanced JTD coherence the charge carriers are more localized and this produces a resistivity increase.

However, we noticed that above P*, the coherence of the Jahn-Teller distortion and bond length only changes slightly with pressure. This is in contrast to the strong pressure dependence of $T_{MI}$, the resistivity and localization length at high pressures. From the structural parameters, it seems that only the Mn-O1-Mn bond angle, which characterizes the tilting of the $MnO_6$ octahedra, changes with pressure above P*. With the $MnO_6$ octahedra more tilted under pressure, the overlap of the $O^{2-}$ 2p orbital and the $e_g$ $d_{3z^2-r^2}$ decreases and the charge carriers are more localized which can be observed in the localization length evolution as a function of pressure (Fig. 5(b)).

It is noticed that the pressure dependence of $T_{MI}$ of our sample above P* is similar to that of the Yttrium doping $La_{1-x-y}Y_yCa_xMnO_3$ system, in which with Y concentration increase $T_{MI}$ decreases and resistivity below $T_{MI}$ increases monotonically.[41, 42, 43] This is ascribed to the $MnO_6$ octahedra buckling. In this system, ferromagnetically correlated clusters or magnetic polarons exist in the paramagnetic insulating phase and applied external magnetic field and spin exchange interaction can affect the localization or magnetic polaron size.[42] Resistance measurements under pressure in magnetic field may help to verify this picture. By comparing these measurements with the pressure dependence of the localization length, one could conclude that with the local structure transformation, the spin state also is changed.

## IV. SUMMARY

High-pressure effects on the resistivity and structure of the CMR material $La_{0.60}Y_{0.07}Ca_{0.33}MnO_3$ have been studied in the pressure range of 1 atm to ~7 GPa. It was found that pressure enhances the ferromagnetic metallic phase and suppresses the resistivity in the measured temperature range below ~3.8 GPa. Above ~3.8 GPa, the resistivity increases and the low temperature ferromagnetic metallic state is suppressed with pressure increase. Structural measurements at room temperature indicate that a structural transformation occurs at ~3.8 GPa consisting of a distortion of the $MnO_6$ octahedra. Above ~3.8 GPa, the buckling of $MnO_6$ octahedra increases with pressure increase. Based on model fits we suggest that the structural changes under pressure leads to localization length or the magnetic cluster (magnetic polaron) size increase at low pressure and decrease at pressures above ~3.8 GPa.

# ACKNOWLEDGMENTS

The high pressure X-ray diffraction measurements were performed at beamline X17B1, NSLS, Brookhaven National Laboratory which is supported by US Department of Energy contract DE-AC02-76CH00016. The authors would also like to thank Dr. Jingzhu Hu at X17C, NSLS for her kind help on the pressure calibration for X-ray diffraction. We are indebted to Prof. John J. Neumeier, at Department of Physics, Montana State University for reviewing the manuscript and giving very useful suggestions. This work is supported by National Science Foundation Career Grant DMR-9733862 and by DMR-0209243.

# Captions

FIG. 1. X-ray diffraction pattern at room temperature and ambient pressure with magnetization measurement shown in the inset (field cooled and zero field cooled).

FIG. 2.  In panel (a) and (b), representative diffraction data at ambient and 5.9 GPa pressure are shown. The shaded regions (not used in the fits) correspond mainly to diffraction from the steel gasket.

FIG. 3. Temperature dependent resistivity curves of $La_{0.60}Y_{0.07}Ca_{0.33}MnO_3$ for varying pressure. Note the shifts in the resistivity peak and change in the amplitude of the resistivity with pressure.

FIG. 4. (a) Pressure dependence of $T_{MI}$.  The metal insulator transition temperature reaches a maximum near 3.8 GPa then decreases rapidly. The solid line is a 3rd order polynomial fit with the coefficient errors in brackets; (b) Pressure dependence of the peak width of the metal-insulator transition. The solid line is a guide to eyes.

FIG. 5. Fit of resistance data with VRH magnetic localization model. (a) plots of the data in the paramagnetic insulating range far from the transition temperature, (the symbolic types are the same as that in FIG. 2); (b) Localization length evaluated with Viret *et al.*[34]s model. The solid line is a guide to eye.

FIG. 6. Pressure dependence of structure parameters for room temperature X-ray diffraction measurements. (a) the unit cell volume; (b) the Mn-O bond lengths of the "ab-plane" Mn-O2 bonds (up and down solid triangles) and "c-axis" Mn-O1 bond (empty squares); (c) the "ab-plane" Mn-O2-Mn (empty squares) and "c-axis" Mn-O1-Mn (solid squares) bond angles; (d) the coherent Jahn-Teller distortion parameter, defined as $\delta_{JT} = \sqrt{\frac{1}{N}\sum(R_{Mn-O} - <R_{Mn-O}>)^2}$. Note that the distortion of $MnO_6$ octahedra reaches a maximum at high pressure.

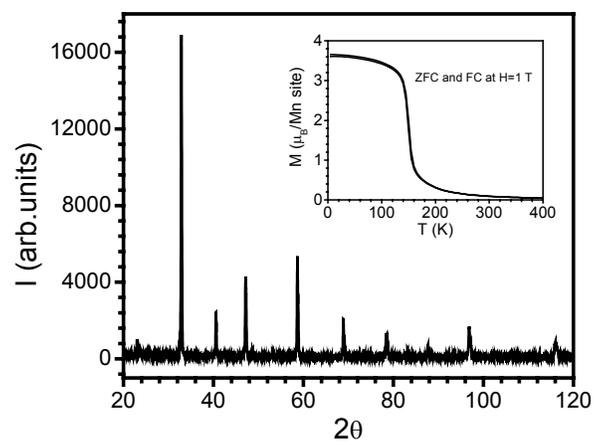

Cui-FIG. 1

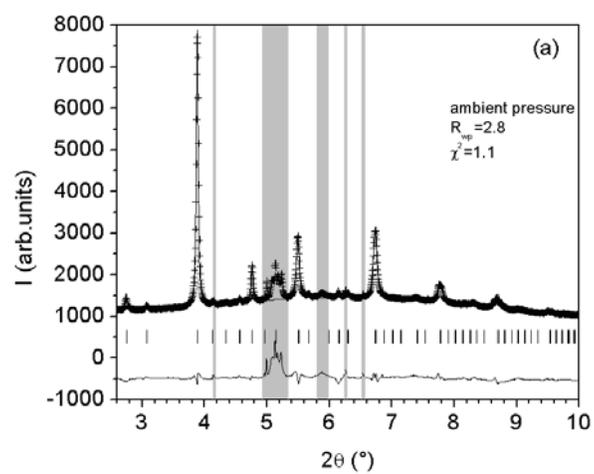

Cui-FIG 2(a)

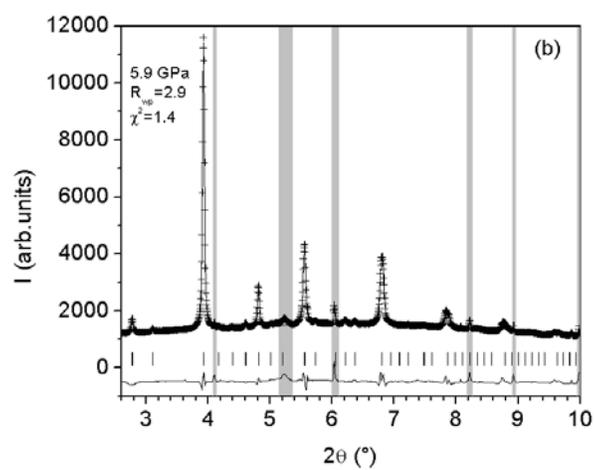

Cui-FIG 2(b)

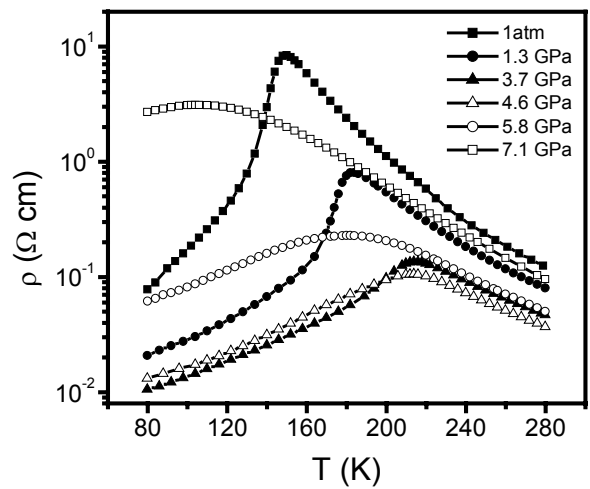

Cui-FIG 3

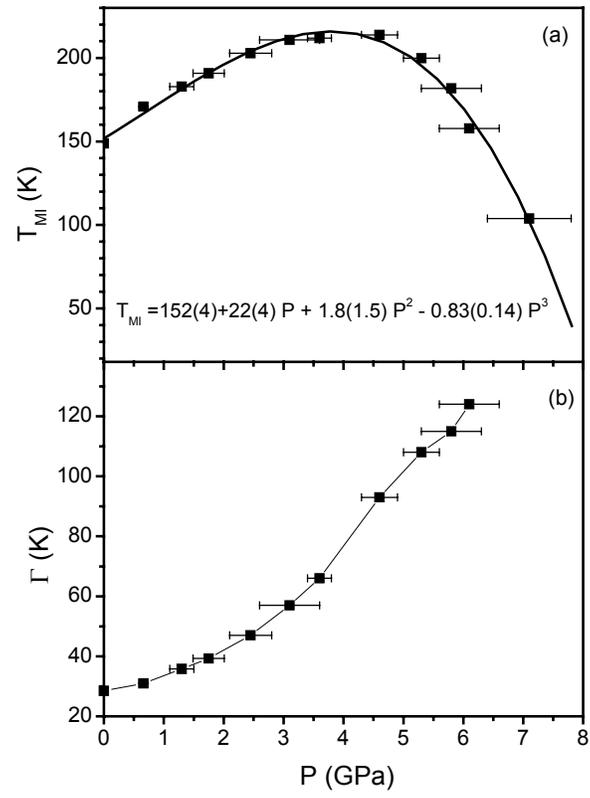

$T_{MI} = 152(4) + 22(4) P + 1.8(1.5) P^2 - 0.83(0.14) P^3$

Cui-FIG. 4

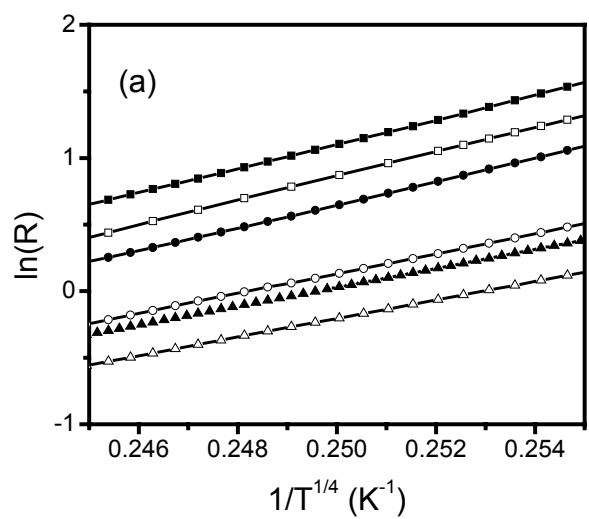

Cui-FIG. 5(a)

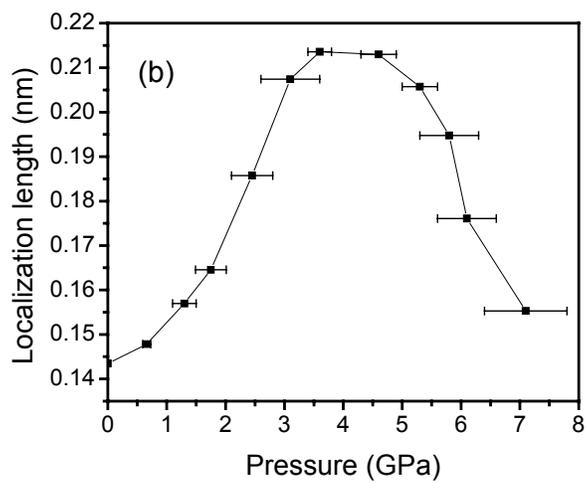

Cui-FIG. 5(b)

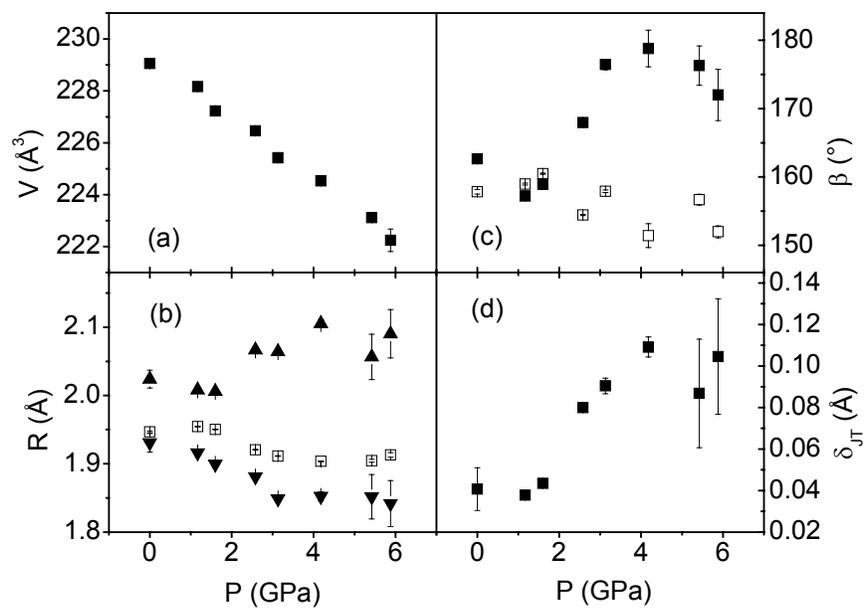

CUI-FIG. 6